\documentclass[showpacs]{revtex4}

\usepackage{amsmath, amsthm, amssymb}
\usepackage{graphicx}

\newcommand{\de}[1]{\left( #1 \right)}
\newcommand{\De}[1]{\left[ #1 \right]}
\newcommand{\DE}[1]{\left\{ #1 \right\}}

\newcommand{\ket}[1]{\left| #1 \right\rangle}
\newcommand{\bra}[1]{\left\langle #1 \right|}

\newcommand{\abs}[1]{\left| #1 \right|}

\newcommand{\tr}{\mathrm{Tr}}

\newcommand{\eg}{{\it{e.g.}} }
\newcommand{\ie}{{\it{i.e.}} }
\newcommand{\etal}{{\it{et al.}}}

\begin{document}
\title{The Geometry of Entanglement Sudden Death}
\author{Marcelo O. Terra Cunha}
\affiliation{Departamento de Matem\'atica, Universidade Federal de Minas Gerais, CP 702, Belo Horizonte, 30123-970, Brazil}
\affiliation{School of Physics and Astronomy, University of Leeds, Leeds, LS2 9JT, UK}
\date{\today}

\begin{abstract}
In open quantum systems, entanglement can vanish faster than coherence. This phenomenon is usually called sudden death of entanglement. In this paper sudden death of entanglement is discussed from a geometrical point of view, in the context of two qubits. A classification of possible scenarios is presented, with important known examples classified. Sugestions of theoretical and experimental examples are given as well as large dimensional and multipartite versions of the effect are briefly discussed.
\pacs{03.65.Ud, 03.65.Yz, 02.40.Ft}
\end{abstract}
\maketitle

\section{Introduction}
Coherence is a central theme in quantum physics. In Feynman's words on quantum theory \cite{Fey}, it is {\emph{the only mystery}}. A good definition of coherence depends on two things \cite{Spek}: the state of the system, $\rho$, and the alternatives under consideration, usually attached to different eigenvalues of an observable $A$. Whenever $\rho$ is written in a basis, the non-diagonal elements of $\rho$ are called {\emph{coherences}} (with respect to that basis) since they can originate the oscillating patterns which characterize interference. Curiously, measurements of $A$ can not show the effects of such coherences. The mean value of $A$ is simply the wheighted sum of its proper values. However, observables which do not commute with $A$ can reveal them. Naturally, when $\rho$ is diagonalized, there is no relevant coherence with respect to that basis. A state which is diagonal in whatever basis can show no coherence at all. In finite dimensions this is the case of the totally mixed state, $\rho _i = \frac{1}{d}I$, $I$ denoting the identity operator and $d$ the dimension of the state space. In this sense, $\rho _i$ can also be called {\emph{the}} incoherent state.

Quantum systems are not isolated. They naturally interact with their environment. When such interaction is taken into account, a {\emph{decoherence}} process usually appears \cite{deco}. In this context, decoherence can be understood as the vanishing of the non-diagonal terms of the density matrix $\rho$, which usually occurs as an exponential decay with a timescale $T_2$ much smaller then the usual damping time $T_1$. Decoherence is even seen as the bridge from quantum mechanics to the classical world, in washing out the interference terms.

In the last decades, a special consequence of ``the only mystery'' has gained central status in quantum mechanics: {\emph{entanglement}}. A good definition of entanglement also depends on two things \cite{single}: the state of the system and a division of such system in meaningfull parties \cite{TPS}. The most paradigmatic context is the bipartite scenario, when the parts are attached to remote experimentalists, Alice and Bob. The state space then takes the form of a tensor product $H = H_A \otimes H_B$. Observables $A \otimes I$ and $I \otimes B$ are {\emph{local}}. States of the form $\rho _A \otimes \rho _B$ show no correlation at all, in the sense that local observables are statistically independent. States of the form
\begin{equation}
\label{sep}
\rho = \sum _i p_i \rho_{Ai} \otimes \rho_{Bi},
\end{equation}
where $p_i$ is a probability distribution, are called {\emph{classically correlated}} or {\emph{separable}} \cite{Wer}. The results of measurements of local observables on separable states can be strongly correlated. A good example, in spin notation, is the state $\rho_s = \frac{1}{2} \de{\ket{\uparrow\uparrow}\bra{\uparrow\uparrow} + \ket{\downarrow\downarrow}\bra{\downarrow\downarrow}}$. The measurements of the $z$-component of spin in both parts are perfectly correlated. But this correlation is classical. A bipartite state is entangled if it can not be written as Eq.~\eqref{sep}.

Recently, interesting effects were recognized in the open system treatment of entangled-state dynamics. In particular, if two qubits are subjected to very natural reservoirs, their entanglement can vanish in finite time, even when coherence only vanishes asymptotically \cite{ZHHH,Diosi,DH,YE02,YE03,YE04,YE06}. This effect was latter called {\emph{Sudden Death of Entanglement}} \cite{SudDet}. It is conjectured to be a quite generic result. A proposal for its experimental detection was already made \cite{San} and implemented \cite{Rio}. However, it is yet considered as a surprising effect, waiting for a simple geometric explanation \cite{YuEbNeg}.

In this article we give such a geometric interpretation to the sudden death of entanglement. This interpretation allows for the explanation of the necessary and sufficient conditions for it to happen. For simplicity, we concentrate in the two qubit scenario, despite the fact that large dimensions and large number of parties can bring new interesting developements \cite{Chile}.

\section{Two Qubits}

Entanglement can be viewed as a consequence of non-local coherence. In fact, a state like $\ket{\Phi_+} = \frac{1}{\sqrt{2}}\DE{\ket{\uparrow\uparrow} + \ket{\downarrow\downarrow}}$ exhibits coherence with respect to the nonlocal obervable $S_{zAB} = S_{zA}\otimes I + I\otimes S_{zB}$. This example is very illustrative, since its entanglement can vanish in a finite time or only asymptotically, depending on the reservoir it is subjected to. For a sudden death situation, one can use independent spontaneous decay with positive temperature \cite{YE04}, while an asymptotic regime is obtained for a dephasing noise \cite{YE03} or for independent spontaneous decay at null temperature \cite{San}.

In this work, the separation between Alice and Bob is fixed (in the sense that we only use the canonical tensor product structure \cite{TPS}), and coherence will be studied with respect to the usual computational basis. For convenience, we will adopt a notation similar to Ref.~\cite{YE02}, by using as a basis
\begin{equation}
\label{basis}
 \begin{array}{cc}
 \ket{1} = \ket{\uparrow\uparrow}, &\ket{2} = \ket{\uparrow\downarrow},\\
 \ket{3} = \ket{\downarrow\uparrow}, &\ket{4} = \ket{\downarrow\downarrow}.
 \end{array}
\end{equation} 
The global density matrix takes the form
\begin{subequations}
\begin{equation}
\label{GDM}
\rho = \De{
 \begin{array}{cccc}
 \rho _{11} &\rho _{12} &\rho _{13} &\rho _{14}\\ 
 \rho _{21} &\rho _{22} &\rho _{23} &\rho _{24}\\ 
 \rho _{31} &\rho _{32} &\rho _{33} &\rho _{34}\\ 
 \rho _{41} &\rho _{42} &\rho _{43} &\rho _{44}
 \end{array}
},
\end{equation}
and the local (reduced) density matrices are
\begin{eqnarray}
\label{LDM}
 \rho _A &=& \De{
  \begin{array}{cc}
  \rho _{11} + \rho _{22} & \rho _{13} + \rho _{24}\\
  \rho _{31} + \rho _{42} & \rho _{33} + \rho _{44}
  \end{array}
 },\\
\rho _B &=& \De{
  \begin{array}{cc}
  \rho _{11} + \rho _{33} & \rho _{12} + \rho _{34}\\
  \rho _{21} + \rho _{43} & \rho _{22} + \rho _{44}
  \end{array}
 }.
\end{eqnarray} 
\end{subequations}

\begin{subequations}
The first important geometric aspect appears here. Our problem starts on a manifold of (real) dimension $15$ given by the positive operators in ${\mathbb{C}}^4$ with unit trace. If we want system $A$ not to show any coherence with respect to the computational basis, we need to have 
\begin{equation}
\label{cohSA}
\rho _{13} + \rho _{24} = 0,
\end{equation}
while if we want to see no coherence for the computational basis of system $B$ we must obey 
\begin{equation}
\label{cohSB}
\rho _{12} + \rho _{34} = 0.
\end{equation}
As each of these equations are complex, they represent a system of two real equations each. Accordingly, each set of states incoherent with respect to local $S_z$ is a subset of dimension $13$. If we want both local coherences to vanish then we get an even smaller subset, of dimension $11$. 

From now on, we will adopt the strategy of focus on the so called $X$ states (due to the appearance of their density matrices, see Eq.~\eqref{spDM}) \cite{YEX}, defined by
\begin{equation}  
 \rho _{12} = \rho_{13} = \rho_{24} = \rho_{34} = 0.
\end{equation}
This set has real dimension $7$. For any natural volume measure, all those sets of local incoherent states have zero measure (like a line in ${\mathbb{R}}^3$). On the opposite direction, the set of unentangled states is known to have positive measure (in particular, the same dimensionality as the ambient) \cite{Zyc}. In a sense (\eg by random choices) it is much more probable to pick an unentangled state than a locally incoherent state.
\end{subequations}

Explicitly reparameterizing the set, we consider density matrices of the form
\begin{equation}
\label{spDM}
\rho = \De{
 \begin{array}{cccc}
 a&0&0&w\\
 0&b&z&0\\
 0&z^*&c&0\\
 w^*&0&0&d
 \end{array}
},
\end{equation}
which shows no local coherence with respect to the computational basis. Despite being ``small'' in the mathematical sense described above, this set contains some important subsets like the states which can be written as a mixture of Bell states (under the conditions $a=d$, $b=c$, and $w,z \in {\mathbb{R}}$), and Werner states \cite{Wer} ($a=d$, $b=c$, $w=0$, and $z=\frac{1-4b}{2}$). 

Let us discuss a little bit more the notions of non-local coherence and separability inside this set.
The block structure of density matrices of the form \eqref{spDM} (more evident if the basis is re-ordered) strongly simplifies the use of Peres-Horodecki criterion \cite{PH} to test for the existence of entanglement. It implies that $\rho$ is entangled iff $\abs{w}^2 > bc$ or $\abs{z}^2 > ad$. The positivity of $\rho$ implies $\abs{w}^2 \leq ad$ and $\abs{z}^2 \leq bc$, hence only one of the above conditions can be fulfilled. By the same conditions, if one of the populations is null, say $d=0$, then $w=0$ and for any non-null value of $z$ the state is entangled. This is a possible strategy to obtain persistent entanglement \cite{YE02,YE03}. It is interesting to note that the elements $z$ and $w$ are non-local coherences, in the sense that they appear in the mean values of obervables like $S_x \otimes S_y$ or $\ket{\Phi_+}\bra{\Phi_+}$. Sudden death of entanglement \cite{SudDet} can then be characterized as the finite time disentanglement of a state, even when non-local coherences are still present, and usually decay exponentially.

\section{Open System Dynamics}
Now let us treat the two qubits as an open system. There is a large collection of natural reservoirs that can be considered (and an even larger one that can be engineered), but let us just divide them into classes. The first question is if they have a single asymptotic state (in the sense that all initial states tend asymptotically to the same point) or not. A good example of the first case is spontaneous decay, while a dephasing noise is an example of the second.
\subsection{A single asymptotic state}
\begin{figure}[tb]
	\centering
   \begin{tabular}{ccc}
		i)\includegraphics[width = 2.5cm]{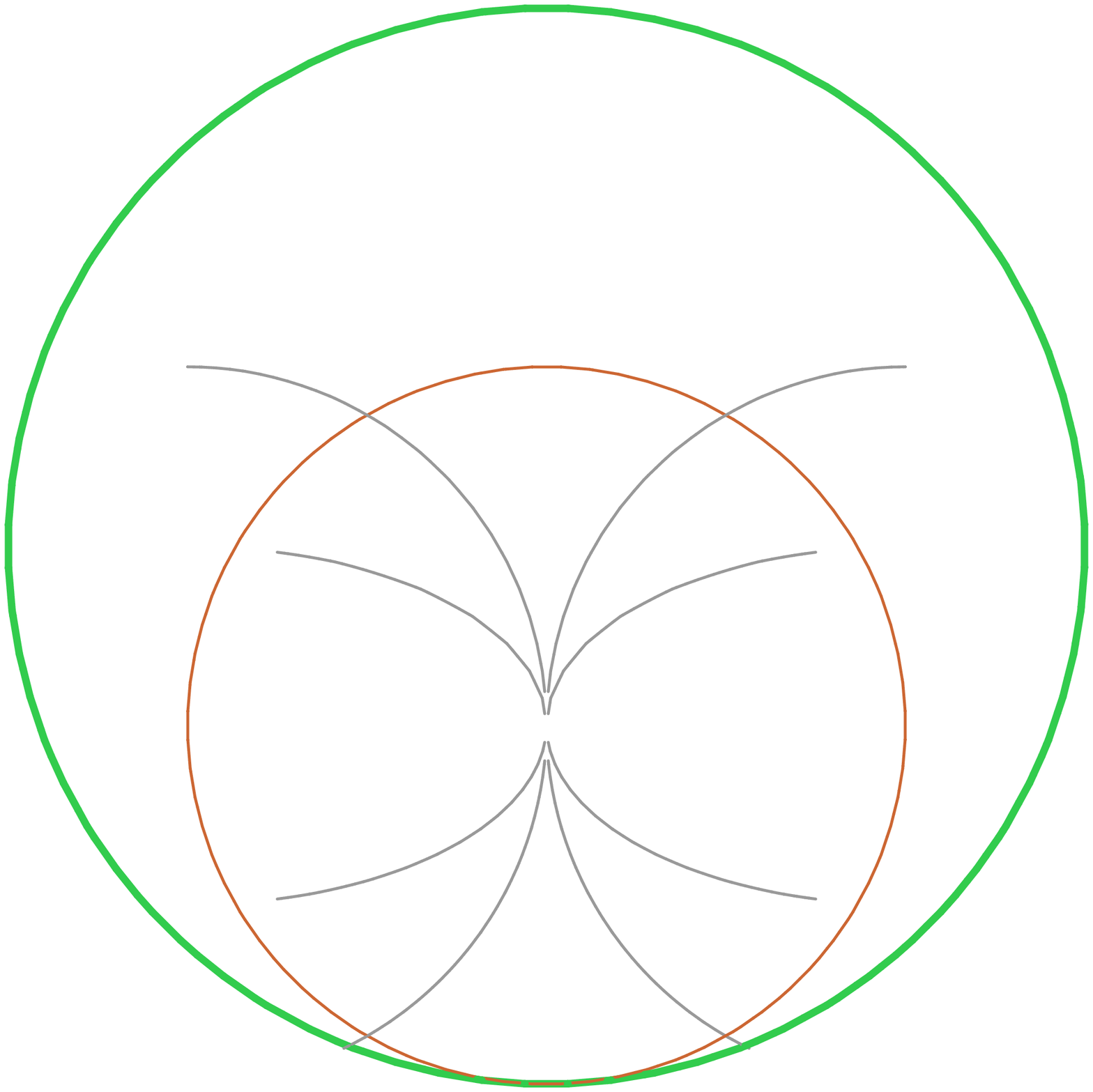}&
		ii)\includegraphics[width = 2.5cm]{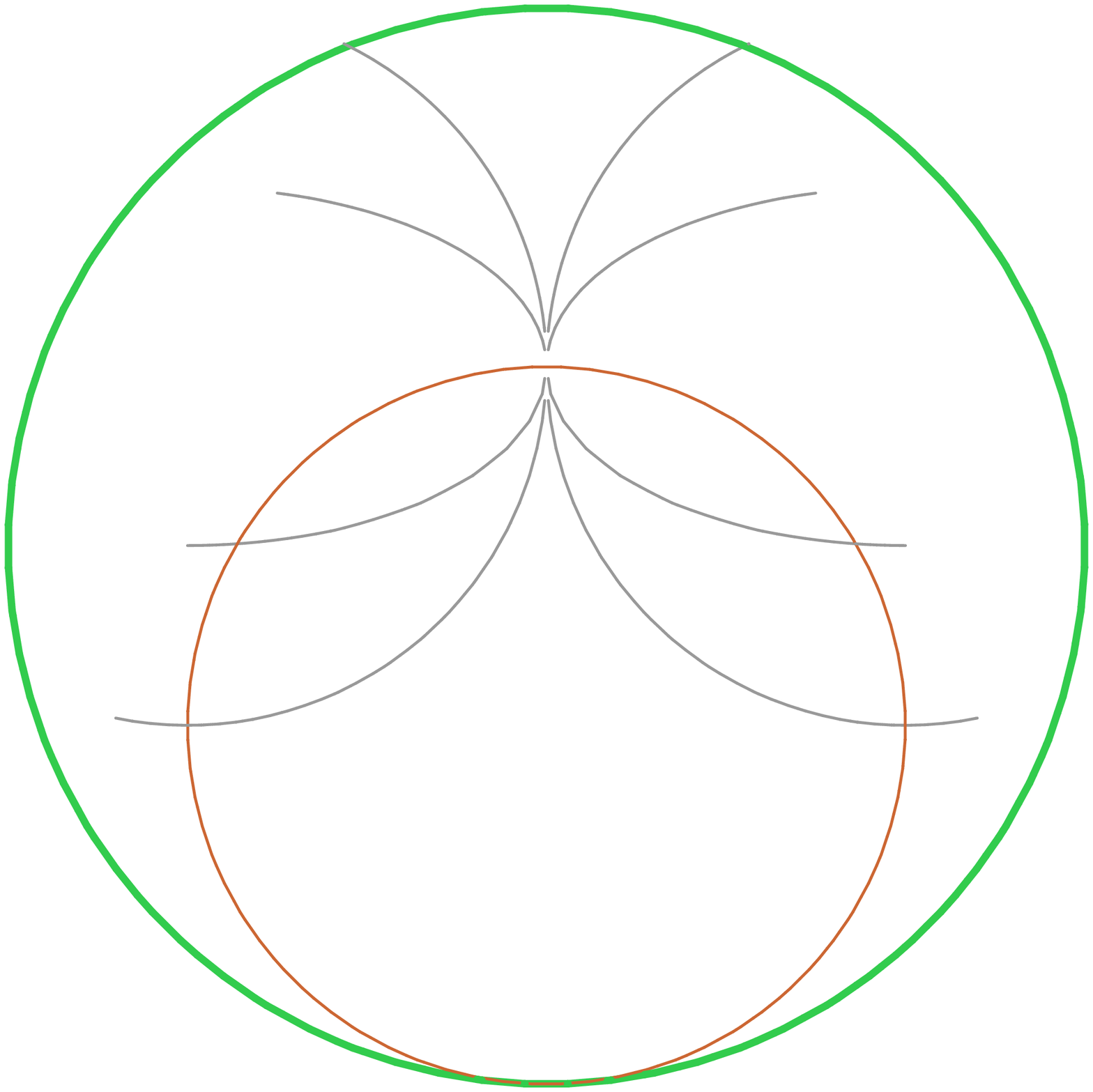}&
		iii)\includegraphics[width = 2.5cm]{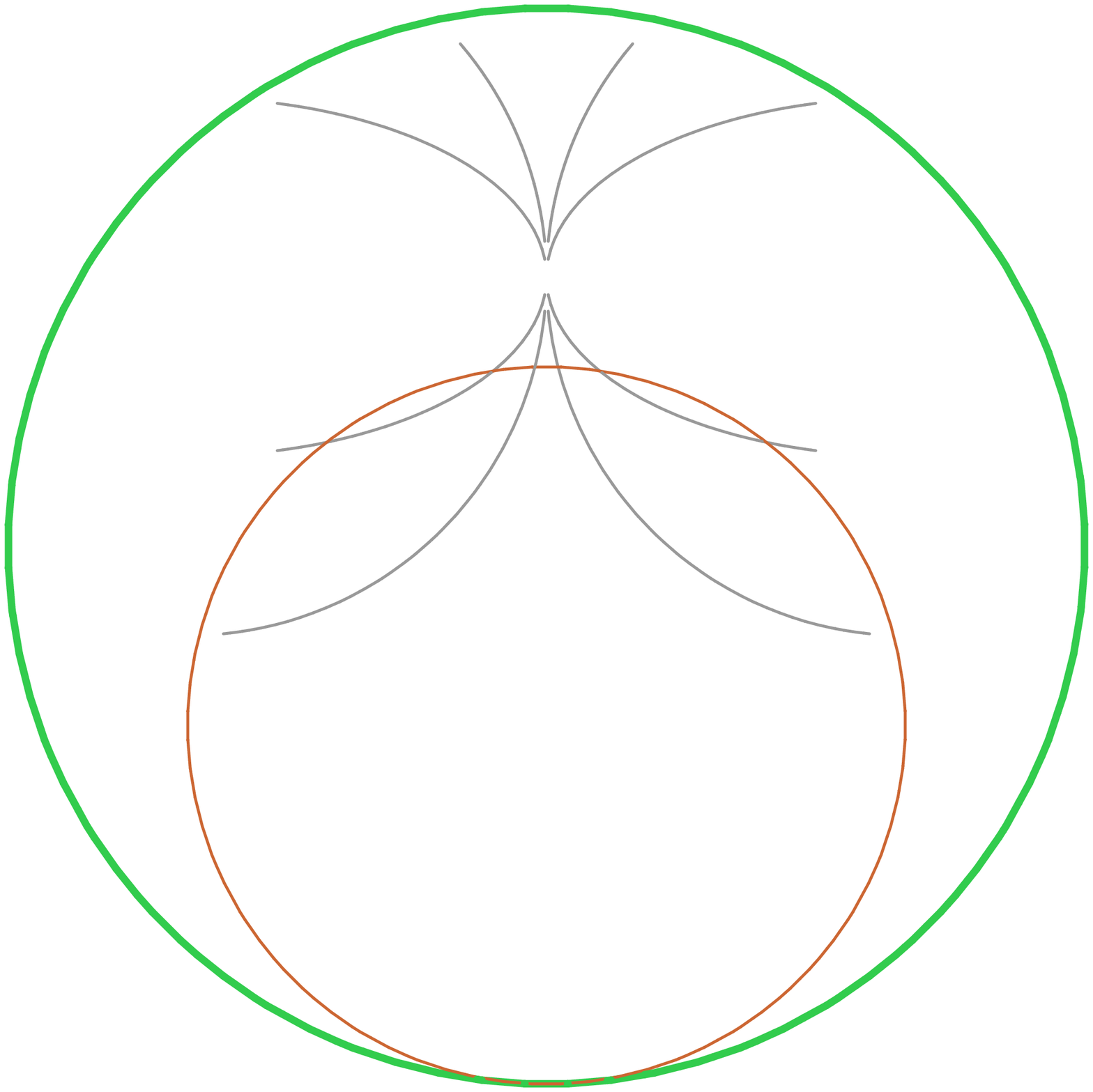}
	 \end{tabular}
	\caption{Schematic pictures of different possible time evolutions with one asymptotic state. In each situation, the outer (green) curve is the boundary of the set of density matrices, the inner (red) curve is the boundary of the set of separable states and the other (grey) curves represent possible time evolutions, all approaching the same asymptotic state. In situation i) the asymptotic state is separable, and entanglement sudden death is the rule; in situation ii) it is still separable, but at the boundary of $S$: both alternatives can coexist, sudden or asymptotic death of entanglement; in situation iii) the equilibrium state is entangled, and the dynamical possibilities include creation of entanglement.}
	\label{oneasymp}
\end{figure}

If the system is subjected to a reservoir with only one asymptotic state, three very diferent situations can happen (see Fig.~\ref{oneasymp}): i) the asymptotic state can be an internal point of the separable states set (\eg for random noise or finite temperature thermal reservoirs for independent qubits); ii) the asymptotic state can be separable but at the boundary of such set (\eg zero temperature spontaneous decay of two independent qubits); or iii) the asymptotic state can be entangled (\eg low temperature reservoirs for some interacting two-qubit Hamiltonians or collective decay of two otherwise independent qubits).

In situation i), sudden death of entanglement is the rule: after a finite amount of time, all initial states will evolve to separable states, by the very definition of an asymptotic interior point and the continuity of time evolution. The reader must appreciate the importance of the reservoirs with this characteristic (one ``deep separable'' asymptotic state) in turning entanglement a generally fragile property of quantum states. The other two situations are richer and deserve more comments.

In situation ii), any neighbourhood of the limit point contains both, entangled and separable states. This is the natural arena for the curious dynamics when some initial states exhibit sudden death of entanglement, while others do not. It is worth noting that usually entanglement discussions are made {\it{modulo}} local unitary operations, since they alone cannot affect the entanglement of the state. However, together with a reservoir which is not invariant under such local unitarities, they can affect the time evolution of entanglement, including the dramatic change from asymptotic separation to finite time disentanglement, and vice-versa! This is well described by the example of states given by Eq.~\eqref{spDM} subjected to independent spontaneous decay at zero temperature. If the ground state is $\ket{4}$, then states with $\abs{w}^2 > ad$ and inverted population $a > d$ show sudden death of entanglement, while for $a \leq d$ the separability is only asymptotic \cite{San}. For this family of states, population inversion is necessary for sudden death to occur. However, this property can be changed by local unitarities (\eg $\ket{\uparrow} \mapsto \ket{\downarrow}$, $\ket{\downarrow} \mapsto -\ket{\uparrow}$).

Situation iii) is not so rare as it can first appear. A generic Hamiltonian for a two-qubit system has entangled ground state \footnote{Since a generic pure state of a composed system is entangled.}. Consequently, null temperature spontaneous decay of interacting qubits is a good example. In such a situation, the asymptotic behaviour shows entanglement, since all states in small neighbourhoods of entangled states are also entangled \footnote{Mathematically, the set of entangled states $E$ is open in the topology induced by the set of all states $D$.}. In this dynamics, after a finite time, any initial state will show entanglement. One could even describe it as a ``sudden birth of entanglement''.

Up to now, we focused on the long time behaviour. Transient evolution can show interesting phenomenology, also. Sudden death of entanglement can be followed by its ``sudden resurrection'' as in Refs.~\cite{ZHHH,FT}. The situation resembles that of different entangled phases on thermal equilibrium states \cite{Osenda,Dow}, and one can even ask whether there are limits in the ``number of lifes'' that entanglement can have, here represented by the number of times that the state trajectory crosses the boundary between entangled and separable states.

\subsection{More than one asymptotic state}
Without extra assumptions, open systems which do not have a unique asymptotic state can be of many types. Time evolving noise is a good example. If we agree on considering only ``static reservoirs,'' then the preservation of convex combinations implies that if there is more than one asymptotic point, all the convex combinations of them are also asymptotic points. This case is well exemplified by phase reservoirs (\ie dephasing noise), when populations are fixed (in the basis defined by the reservoir) and (usually) coherences drop exponentially.

\begin{figure}[tb]
	\centering
   \begin{tabular}{cccc}
		i)\includegraphics[width=2.5cm]{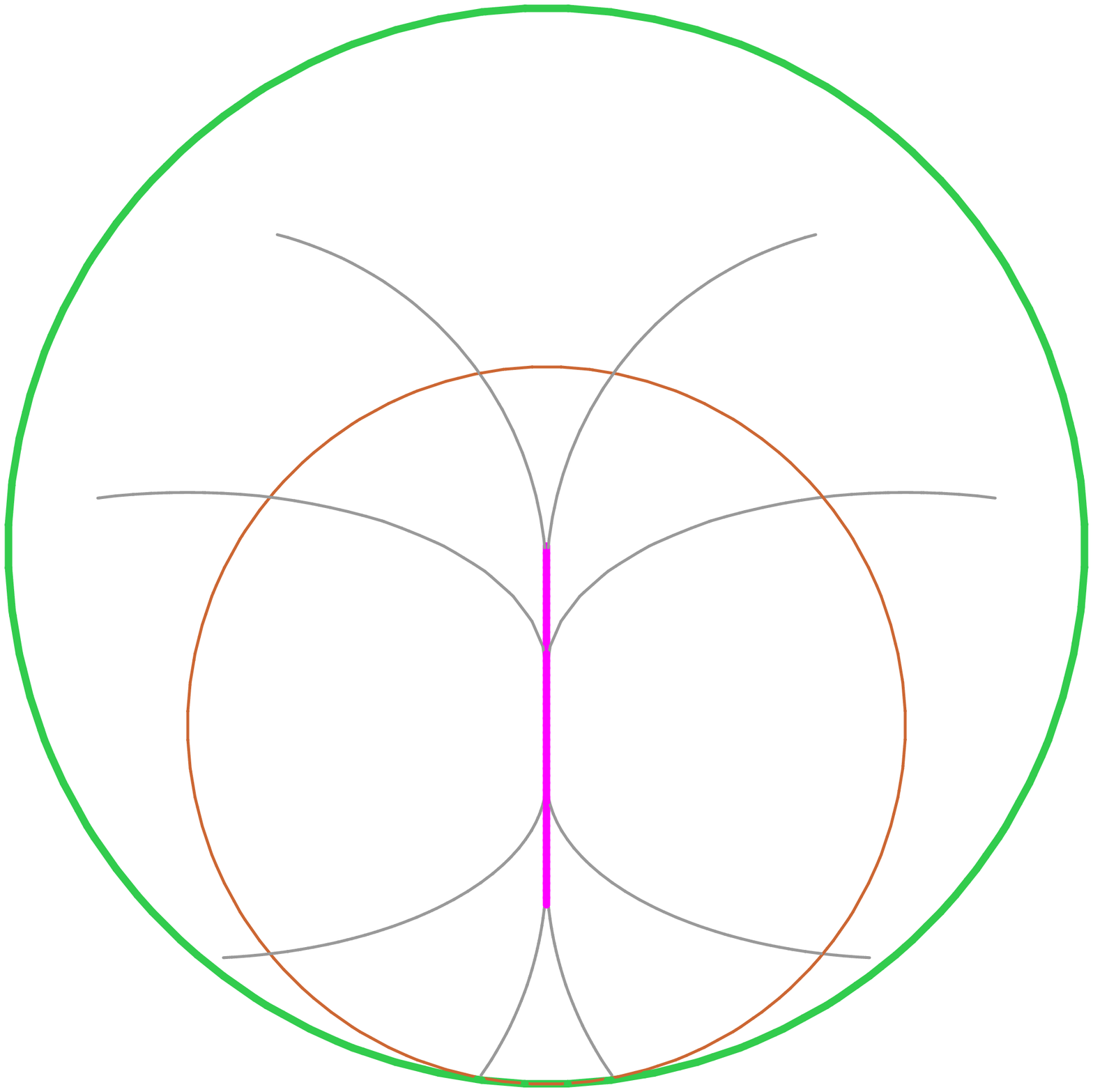}&
		ii)\includegraphics[width=2.5cm]{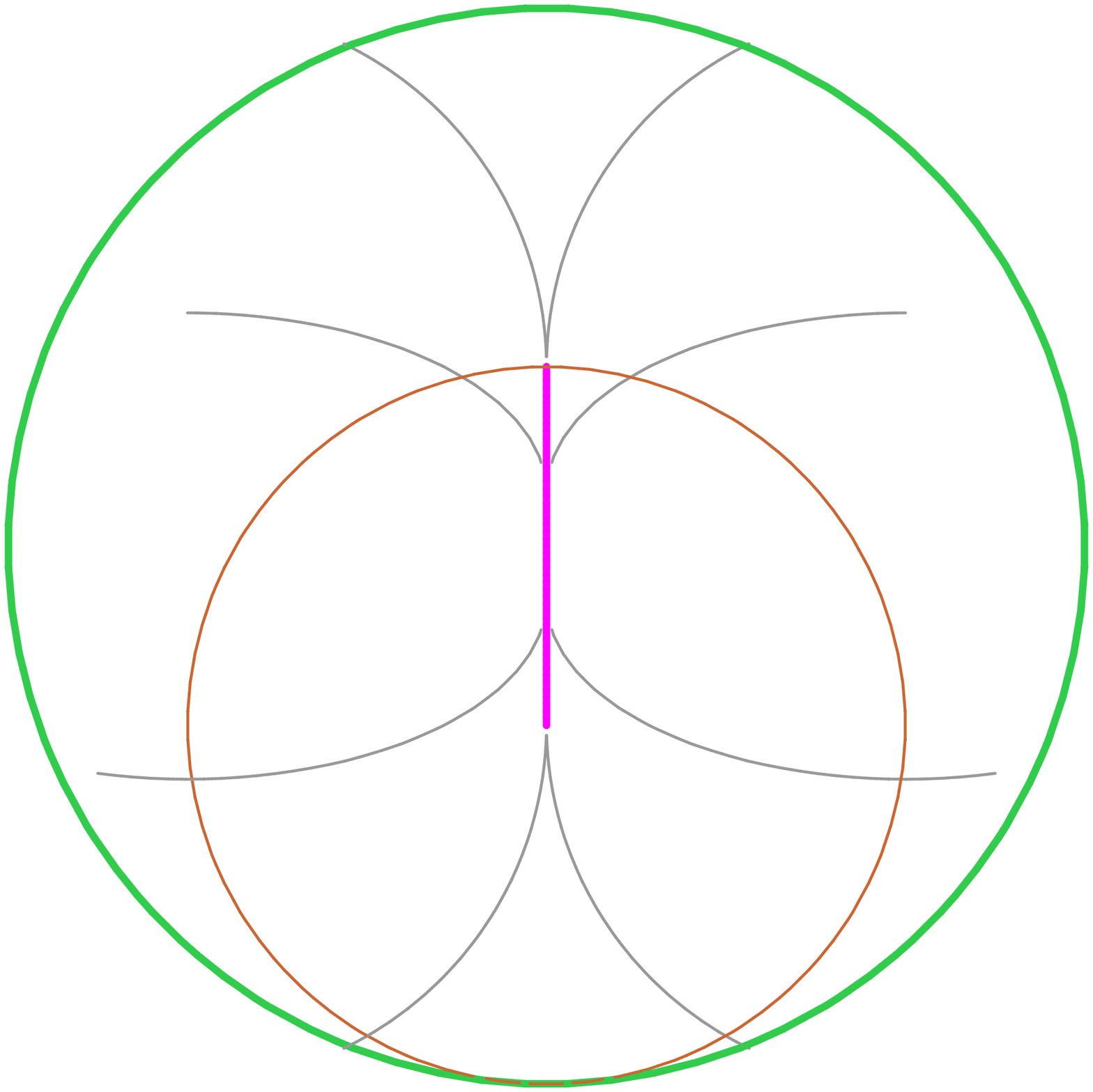}&
		iii)\includegraphics[width=2.5cm]{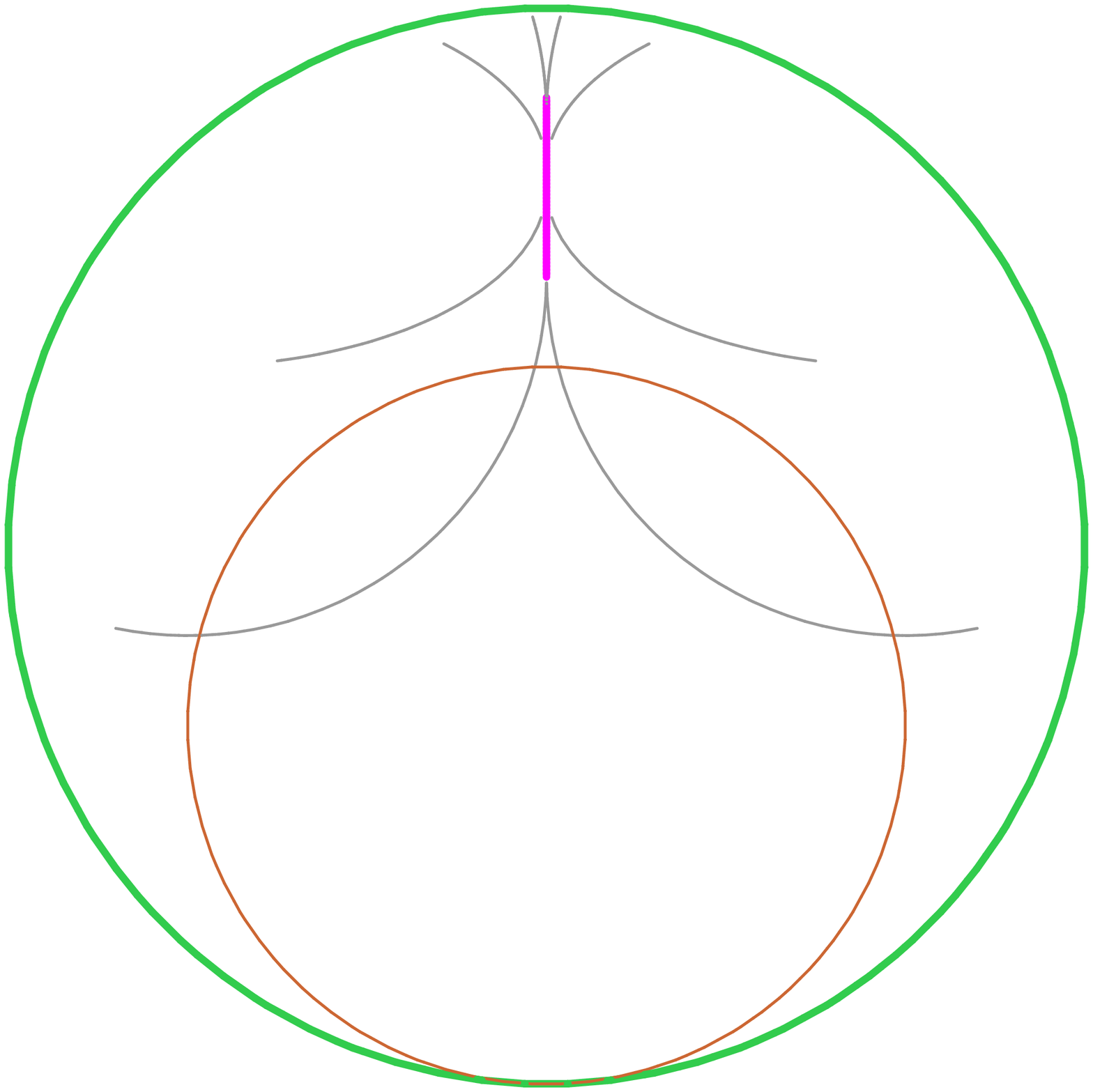}&
		iv)\includegraphics[width=2.5cm]{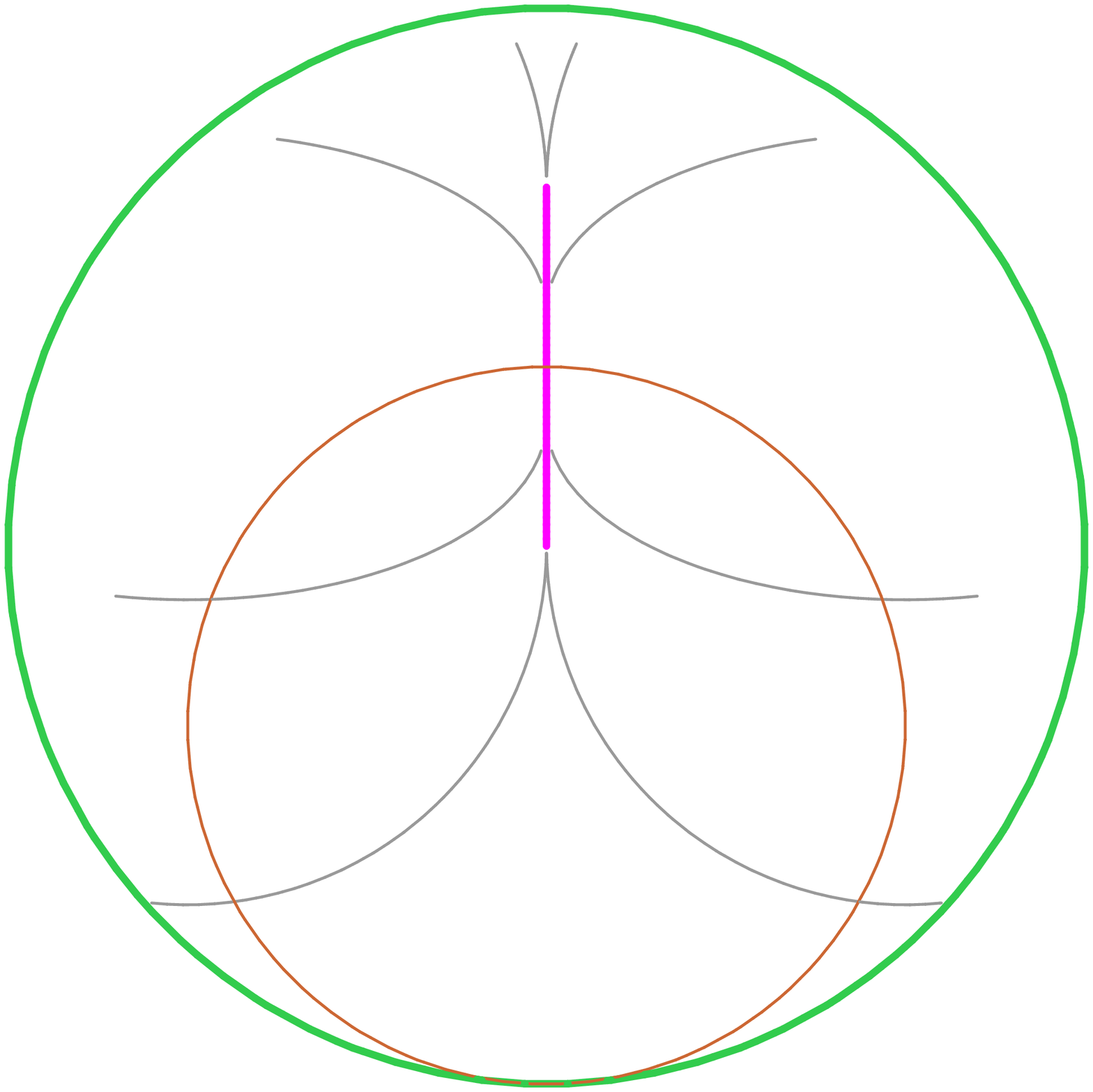}
	 \end{tabular}
	\caption{Possible situations for multiple asymptotic states. The central (magenta) line represents the set $R$ of asymptotic states. Situations i) to iii) are analogous to the previous case. Situation iv) occurs for $R$ containing both, separable and entangled states. Despite its formal difference, its qualitative behaviour is very similar to situation ii) of the firts case.}
	\label{mult}
\end{figure}

Geometrically, the situation now is only a little trickier (see Fig.~\ref{mult}). Let us call $R$ the set of asymptotic states, $S$ the set of separable ones, $S^o$ its interior, and $E$ the set of entangled states. Sets $R$ and $S$ are convex and closed. Their relative position is essential here. If $R \subset S^o$ we have the generalization of the situation i) of the previous case. Again sudden death is the rule in this situation, with some preceding entanglement revivals allowed. Another imediate generalization happens if $R \subset S$, but $R \not\subset S^o$, \ie when there are asymptotic points in the boundary of the separable set. Here sudden death and persistent entanglement can, in principle, coexist, depending on the specific route each initial state takes and whether its asymptotic state be at the boundary $\partial S$ \footnote{Remember that the system has a large set of asymptotic states $R$, but as the dynamic is defined, each initial condition tends to one specific asymptotic state $\rho \in R$.}. One natural example of this geometric configuration is given by two non-interacting qubits subjected to independent dephasing noise \cite{YE03}. In this example, $R$ is given by a tetrahedron with the base states \eqref{basis} on the vertexes, \ie states of the form \eqref{spDM} with $w=z=0$. All the faces of this tetrahedron are at the boundary of $S$, $\partial S$. Those faces are characterized by one null eigenvalue, say $d$. This face can be aproximated by states with $z$ approaching zero, which are non-separable (analogous phenomena happen to the other faces). The more interesting geometric aspect happens with the points we would like to call the interior of the tetrahedron (those with only non-null populations). All these points are interior to $S$, and the dephasing noise allows them to be approached without touching the faces of the tetrahedron (like one can touch a point in the interior of a triangle in a piece of paper without crossing its edges). For this example, sudden death is the rule for general inial states, while asymptotic disentanglement is allowed for states which tend to the faces (or some edges) of the set $R$.

The generalization of situation iii) of the previous case, corresponding to the new situation iii) when $R \subset E$, is also geometrically possible, but the non-convexity of $E$ makes it harder to find dynamically. If one takes two different entangled states, there is no guarantee that their convex combinations remain entangled. The most famous example is for a pair of Bell states: any equal mixture of two Bell states is separable \footnote{Just apply Peres-Horodecki criterion.}. Another important restriction is that $R$ can not contain a whole set of projectors $\ket{\psi}\bra{\psi}$ for any two-dimensional subspace of $H \equiv {\mathbb{C}}^4$ \footnote{The set of separable states corresponds to the Segre embeding ${\mathbb{C}}P^1 \times {\mathbb{C}}P^1 \rightarrow {\mathbb{C}}P^3$, a quadric complex surface, while each two-dimensional subspace of ${\mathbb{C}}^4$ corresponds to a projective line ${\mathbb{C}}P^1 \subset {\mathbb{C}}P^3$. This two sets must intersect, by the Bezout Theorem \cite{Harris}.}. With respect to the dynamics of entanglement, this situation behaves in a way analogous to the single entangled asymptotic point: after a finite amount of time, only entangled states can be found. To obtain Krauss operators for a reservoir like this is yet an open problem. 

The remaining possibility happens when $R$ contains both separable and entangled states. For this situation, all the behaviours previously discussed are, in principle, possible. Let us just discuss one general and interesting example of this class: two non-interacting identical qubits subjected to collective phase noise. In this situation, the asymptotic states are of the form \eqref{spDM} with $w=0$. This is a five dimensional manifold (remember $z$ is complex and $\tr \rho = 1$) containing both separable and entangled states. In this example, initial states like %
\begin{subequations}
\begin{equation}
\label{Psiw}
\rho = \De{
 \begin{array}{cccc}
 a&0&0&w\\
 0&0&0&0\\
 0&0&0&0\\
 w^*&0&0&d
 \end{array}
}
\end{equation}
exhibit asymptotic decay of entanglement, while
\begin{equation}
\label{Psiwbc}
\rho = \De{
 \begin{array}{cccc}
 a&0&0&w\\
 0&b&0&0\\
 0&0&c&0\\
 w^*&0&0&d
 \end{array}
}
\end{equation}
with $\abs{w}^2 > bc > 0$ show sudden death; on the other hand, initially separable states remain separable for all times, while entangled states like
\begin{equation}
\label{Phiz}
\rho = \De{
 \begin{array}{cccc}
 0&0&0&0\\
 0&b&z&0\\
 0&z^*&c&0\\
 0&0&0&0
 \end{array}
}
\end{equation}
remain fixed as a consequence of the existence of a decoherence free subspace \cite{DFS}.
\end{subequations}
\section{Concluding remarks}
To conclude, we affirm that the curious effect of sudden death of entanglement is an example of a large class of interesting behaviours which open quantum system can show. It also brings together a rich geometrical flavour. This geometrical reasoning makes simpler the decision of when entanglement should be destroyed in finite time, when it can even be created by damping, and when qualitatively different time evolutions can coexist in the same system. A natural question which appears from this work is how to give examples, in theory and in laboratory, for each different possible geometrical situation. Another is about the existence of a maximum number of possible ``entanglement resurrections'' under some natural restrictions on the reservoirs. Generalizations are also in order: large dimensional bipartite systems can exhibit new features with respect to sudden death of entanglement and its geometry. For example, has bound entanglement any special behaviour with respect to sudden death? Another natural generalization deals with multipartite entanglement. It is clear from the geometrical picture that the effect can take place in such context, but as now different kinds of entanglement appear, the classification scheme becomes much trickier. 

This work was originated in discussions with Dr. Marcelo Fran\c ca Santos and Daniel Cavalcanti. The author thanks their suggestions, as well as comments from Dr. Karol \.Zyczkowski and Raphael Drumond. Financial support from CNPq, Fapemig, and PRPq-UFMG are kindly acknowledged. This work is part of the Millenium Project for Quantum Information, CNPq.


\end{document}